\documentclass[10pt,a4paper]{iopart}
\usepackage{iopams,graphicx}
\usepackage[english]{babel}

\newcommand{\identity}{\leavevmode\hbox{\small1\normalsize\kern-.33em1}} % Identity operator.
\newcommand{\difx}[1]{\rmd #1\;}                                         % Roman dee, then argument, then space.

\begin{document}

\title[A QWEI for the Dirac field in two-dimensional flat spacetime]{A quantum weak energy inequality for the Dirac
field in two-dimensional flat spacetime}

\author{S P Dawson}

\address{Department of Mathematics, University of York, Heslington, York YO10 5DD, UK}

\ead{spdawson@gmail.com}

\begin{abstract}
Fewster and Mistry have given an explicit, non-optimal quantum weak energy inequality that constrains the smeared energy
density of Dirac fields in Minkowski spacetime. Here, their argument is adapted to the case of flat, two-dimensional
spacetime. The non-optimal bound thereby obtained has the same order of magnitude, in the limit of zero mass, as the
optimal bound of Vollick. In contrast with Vollick's bound, the bound presented here holds for all (non-negative) values
of the field mass $m$.
\end{abstract}

% PACS numbers for: theory of quantized fields, field theory.
\pacs{03.70.+k, 11.10.-z}

\section{Introduction}

Every observed classical matter field obeys certain \emph{energy conditions} \cite{Hawking/Ellis}; these constrain the
local energy density of the field to be everywhere non-negative. The situation in quantum field theory is very
different: it is well known that the energy density of a quantum field may have a negative expectation value at a point
\cite{Epstein/Glaser/Jaffe}.

Although the classical energy conditions do not hold directly for quantum fields, the correspondence principle gives
good reason to suppose that some analogous quantum energy conditions might be found. Indeed, were this not the case,
then the resultant `exotic' matter fields could potentially be used for a number of alarming subversions of physical law
--- these include large-scale violations of the second law of thermodynamics \cite{Ford 1978}, the creation of
traversable wormholes \cite{Ford/Roman 1996}, and the potential for faster-than-light travel using `warp drive'
technologies \cite{Pfenning/Ford warp, Alcubierre, Everett/Roman, Krasnikov}. Fortunately, it transpires that quantum
field theory does contain a mechanism that prevents unconstrained negative energy densities. This mechanism is made
manifest in what shall be known as \emph{quantum inequalities} or \emph{quantum weak energy inequalities} (QWEIs) ---
this latter description emphasizing the parallel with the weak energy condition of classical field theory.

The derivation and study of QWEIs was initiated by Ford and co-workers \cite{Ford 1978, Ford 1991, Ford/Roman 1995,
Ford/Roman 1997}, culminating in a bound for the scalar field in static spacetimes \cite{Pfenning/Ford static}. These
early bounds were generalized in the work of Fewster and collaborators \cite{Fewster/Eveson, Fewster/Teo}, culminating
in the most general result for the scalar field: a bound holding in any globally-hyperbolic spacetime \cite{Fewster
general}. In addition to the scalar field, QWEIs have been found for the Maxwell and Proca fields (both spin 1); the
earliest results for spin 1 fields were given in \cite{Ford/Roman 1997}, and later generalised in the findings of
\cite{Pfenning, Fewster/Pfenning}. QWEIs have also been obtained for the Rarita-Schwinger (spin $\frac{3}{2}$) field
\cite{Yu/Wu}.

The fundamental object of study is the \emph{smeared energy density} $T_{g^2}$; this is a weighted average of the energy
density, defined as
\begin{equation}\label{definition of T_f}
  T_{g^2}:=\int\difx{t}:T_{00}\left(t,x_0\right):g\left(t\right)^2.
\end{equation}
For convenience, and with no loss of generality, the real-valued \emph{smearing function} $g\in
C^\infty_0\left(\mathbb{R}\right)$ is assumed to have unit $L^1$ norm. A QWEI is a state-independent lower bound on the
expectation value of $T_{g^2}$, and is said to be \emph{optimal} if the lower bound is an infimum over the class of
states which are `physically meaningful' in some appropriate sense. Here, this class will consist of those states in the
Fock space of the Minkowski vacuum for which the manipulations described are valid. Although this class will not be
explicitly delineated here, it is expected to include all states for which contributions from high momenta or high
particle number are damped (e.g. exponentially). The space of these `physically meaningful' states is dense in Fock
space. Fully rigorous treatments of QWEIs, such as \cite{Fewster general, Fewster/Verch}, do not need to fix a Hilbert
space representation and can consider the full class of Hadamard states.

Flanagan \cite{Flanagan} gave the first optimal bound, for the \emph{massless} scalar field in two-dimensional flat
spacetime. Vollick \cite{Vollick} subsequently showed that Flanagan's bound also applies to, and is optimal for,
massless Dirac fields on two-dimensional flat spacetime --- this was the first quantum inequality for the Dirac field.
Vollick's bound holds only for static observers in static spacetimes; Flanagan \cite{Flanagan 2002} succeeded in
removing this restriction. In Minkowski spacetime, the Flanagan-Vollick bound has the form
\begin{displaymath}
  \langle T_{g^2}\rangle_\psi
    \geq-\frac{1}{6\pi}\int_{-\infty}^\infty\difx{t}\left|\frac{\rmd g}{\rmd t}\right|^2,
\end{displaymath}
where $T_{g^2}$ is defined by (\ref{definition of T_f}). The arguments used to derive this optimal bound rely heavily on
certain conformal properties of massless fields in two dimensions, and do not generalize directly to non-zero masses or
higher-dimensional spacetimes. A general treatment of two-dimensional conformal field theories has been given in
\cite{Fewster/Hollands}; this work makes the important generalization to the case of interacting field theories.

Fewster and Verch \cite{Fewster/Verch} have shown that a bound exists for the Dirac field in any four-dimensional
globally-hyperbolic spacetime; however, this bound is not given by an explicit formula. Fewster and Mistry
\cite{Fewster/Mistry} have given the first explicit, closed-form bound on the smeared energy density of a Dirac field on
four-dimensional flat spacetime. Here, their argument (which requires only the anticommutation relations and a certain
identity for Fourier transforms) is adapted to the case of two-dimensional flat spacetime, leading to the bound
\begin{equation}\label{the final bound}
  \langle T_{g^2}\rangle_\psi
    \geq-\frac{1}{2\pi^2}\int_m^\infty\difx{u}u^2\left|\widehat{g}\left(u\right)\right|^2
    Q_1^D\left(\frac{u}{m}\right),
    \quad
    m \geq 0.
\end{equation}
Here, $m$ is the field mass, and
\begin{equation}\label{definition of Q_1^D}
  Q_1^D\left(x\right):=\sqrt{1-\frac{1}{x^2}}-\frac{1}{x^2}\ln\left(x+\sqrt{x^2-1}\right),\quad 1\leq x\leq\infty.
\end{equation}
Like the four-dimensional bound of \cite{Fewster/Mistry} (and in contrast to that of \cite{Fewster/Verch}), the bound
given here is explicit, and in closed-form. The bound is easily seen to be non-optimal by consideration of the massless
limit; here, the present bound is weaker by a factor of 3 than the Flanagan-Vollick optimal bound. It is worth
emphasizing that, although it is not optimal, the present bound does apply to the case $m>0$; the Flanagan-Vollick
bound, in contrast, holds strictly for the special case of zero mass.

\section{The Dirac equation on two-dimensional flat spacetime}

Let spacetime be flat and two-dimensional, and let the metric signature be $\left(+-\right)$. Then, choosing a system of
units in which $\hbar=c=1$, a fermion field $\psi$ of mass $m$ satisfies the Dirac equation
\begin{displaymath}
  \left(\rmi\gamma^\mu\partial_\mu-m\right)\psi=0.
\end{displaymath}
The $2\times2$ matrices $\gamma^\mu$ obey
\begin{displaymath}
  \left\{\gamma^\mu,\gamma^\nu\right\}=2\eta^{\mu\nu}\identity,\quad\mu,\nu=0,1.
\end{displaymath}
The field $\psi$ is a two-component spinor. In the sequel, spinor indices will often be suppressed: $\psi$ itself is a
notational shorthand for
\begin{displaymath}
  \psi=\left(\psi_i\right)=\left(\begin{array}{c}\psi_1 \\ \psi_2\end{array}\right).
\end{displaymath}
A good example of the suppression of spinor indices is provided by the spinor product $\left\|\psi\right\|^2$; written
out fully, this is
\begin{displaymath}
  \left\|\psi\right\|^2:=
  \psi^\ast\psi=\left(\psi^\ast_1,\psi^\ast_2\right)\left(\begin{array}{c}\psi_1 \\ \psi_2\end{array}\right)
  =\psi^\ast_1\psi_1+\psi^\ast_2\psi_2=\sum_i\psi^\ast_i\psi_i.
\end{displaymath}

For calculational simplicity, the field will be quantized in a one-dimensional `box' of side length $2L$; the continuum
limit $L\rightarrow\infty$ will be taken later. The field operator $\psi$ can be expanded in terms of creation and
annihilation operators as
\begin{equation}\label{mode expansion}
  \psi\left(t,x\right)=\sum_{k}\left[b_ku_k\rme^{-\rmi\left(\omega_kt-kx\right)}
    +d_k^\ast v_k\rme^{+\rmi\left(\omega_kt-kx\right)}\right],
\end{equation}
where $\left(L/\pi\right)k\in\mathbb{Z}$, and the mode energy $\omega_k$ is defined as
\begin{displaymath}
  \omega_k:=\sqrt{k^2+m^2}.
\end{displaymath}
The basis spinors $u_k$ and $v_k$ obey periodic boundary conditions on $\left[-L,L\right]$, as a consequence of the
`box' quantization. Additionally, $u_k$ and $v_k$ are assumed to be normalized so that
\begin{equation}\label{spinor normalization}
  \left\|u_k\right\|^2=\left\|v_k\right\|^2=\frac{1}{2L}.
\end{equation}
This choice of normalization ensures that both $u_k \rme^{-\rmi\left(\omega_k t-kx\right)}$ and
$v_k\rme^{+\rmi\left(\omega_kt-kx\right)}$ have unit $L^2$ norm on $\left[-L,L\right]$. The creation and annihilation
operators obey the anticommutation relations
\begin{equation}\label{anticommutation relations 1}
  \left\{b_k,b^\ast_{k^\prime}\right\}=\delta_{kk^\prime}\identity
\end{equation}
and
\begin{equation}\label{anticommutation relations 2}
  \left\{d_k,d^\ast_{k^\prime}\right\}=\delta_{kk^\prime}\identity.
\end{equation}
All other anticommutators of creation and annihilation operators vanish identically.

The energy density of the quantized field is obtained by substitution of the mode expansion (\ref{mode expansion}) into
the classical expression
\begin{displaymath}
  T_{00}=\frac{\rmi}{2}\left[\psi^\ast\dot{\psi}-\dot{\psi}^\ast\psi\right]
\end{displaymath}
for the energy density. The result, evaluated at the spatial origin $\left(t,0\right)$, is
\begin{eqnarray}\label{energy density at spatial origin}
  \fl
  :T_{00}\left(t,0\right):  =  \frac{1}{2}\sum_{k,k^\prime}\Bigl\{
    \left(\omega_k+\omega_{k^\prime}\right)\left[
      b^\ast_kb_{k^\prime}u^\ast_ku_{k^\prime}\rme^{-\rmi\left(\omega_{k^\prime}-\omega_k\right)t}
      +d^\ast_{k^\prime}d_{k}v^\ast_kv_{k^\prime}\rme^{-\rmi\left(\omega_k-\omega_{k^\prime}\right)t}
      \right] \nonumber \\
    {} +
    \left(\omega_{k^\prime}-\omega_k\right)\left[
      d_kb_{k^\prime}v^\ast_ku_{k^\prime}\rme^{-\rmi\left(\omega_k+\omega_{k^\prime}\right)t}
      -b^\ast_kd^\ast_{k^\prime}u^\ast_kv_{k^\prime}\rme^{+\rmi\left(\omega_k+\omega_{k^\prime}\right)t}
    \right]
  \Bigr\}.
\end{eqnarray}
The normal-ordered operator has been given, making use of the rule
\begin{equation}\label{normal ordering rule}
  :d_kd^\ast_{k^\prime}:=-d^\ast_{k^\prime}d_k.
\end{equation}

It will have been noticed that the spinors $u_k$ and $v_k$ appearing in the mode expansion (\ref{mode expansion}) carry
labels for the mode momentum only. This is in contrast to the familiar four-dimensional case, where the corresponding
basis spinors are often written $u_k^\alpha$ and $v_k^\alpha$, with the index $\alpha=1,2$ labelling the two independent
spin states. Spin is essentially trivial in two dimensions \cite{Green/Schwarz/Witten}, and so a spin index is not
required.

\section{Derivation of the quantum weak energy inequality}

Writing $f=g^2$, where $g\in C^\infty_0\left(\mathbb{R}\right)$ is real-valued and has unit $L^1$ norm, the smeared
energy density measured by an observer static at the spatial origin is the expectation value of
\begin{displaymath}
  T_f:=\int\difx{t}:T_{00}\left(t,0\right):f\left(t\right).
\end{displaymath}
(The translational invariance of the theory will be employed later, to remove the specialization to the spatial origin.)
What is required is a lower bound on the spectrum of $T_f$. Equation (\ref{energy density at spatial origin}) gives
\begin{eqnarray}\label{to be compared with}
  \fl
  T_f & =  \frac{1}{2}\sum_{k,k^\prime} \Bigl\{
    \left(\omega_k+\omega_{k^\prime}\right)\left[
      b^\ast_kb_{k^\prime}u^\ast_ku_{k^\prime}\widehat{f}\left(\omega_{k^\prime}-\omega_k\right)
      +d^\ast_{k^\prime}d_{k}v^\ast_kv_{k^\prime}\widehat{f}\left(\omega_k-\omega_{k^\prime}\right)
      \right] \nonumber \\
    \fl & {} +
    \left(\omega_{k^\prime}-\omega_k\right)\left[
      d_kb_{k^\prime}v^\ast_ku_{k^\prime}\widehat{f}\left(\omega_k+\omega_{k^\prime}\right)
      -b^\ast_kd^\ast_{k^\prime}u^\ast_kv_{k^\prime}\widehat{f}\left(-\omega_k-\omega_{k^\prime}\right)
    \right]
  \Bigr\},
\end{eqnarray}
where the Fourier transform $\widehat{f}$ of a function $f:\mathbb{R}\rightarrow\mathbb{R}$ is defined as
\begin{displaymath}
  \widehat{f}\left(\omega\right):=\int_{-\infty}^\infty\difx{t}f\left(t\right)\rme^{-\rmi\omega t}.
\end{displaymath}

The first step in the derivation is the definition of a family of spinorial Fock space operators:
\begin{displaymath}
  \fl
  \mathcal{O}_{\mu,i}:=\sum_k\left\{
    \overline{\widehat{g}\left(-\omega_k+\mu\right)}b_k u_{k,i}
    +\overline{\widehat{g}\left(\omega_k+\mu\right)}d^\ast_k v_{k,i}
    \right\}, \quad \mu\in\mathbb{R}, \quad i=1,2.
\end{displaymath}
(Here, in order to avoid confusion with the continuous parameter index $\mu$, the spinor index $i$ is shown explicitly.)
The derivation now proceeds very much as in \cite{Fewster/Mistry}: the technique is to rewrite $T_f$ in terms of
integrals of $\mathcal{O}_{\mu,i}^\ast \mathcal{O}_{\mu,i}$ and $\mathcal{O}_{\mu,i} \mathcal{O}_{\mu,i}^\ast$, plus a
multiple of the identity which emerges from the canonical anticommutation relations. Specifically
\begin{displaymath}
  \fl
  T_f  = \frac{1}{2\pi}\int_0^\infty\difx{\mu}\mu\left(\mathcal{O}^\ast_{\mu,i}\mathcal{O}_{\mu,i}-S_{\mu}\identity\right)
          +\frac{1}{2\pi}\int_{-\infty}^0\difx{\mu}\mu\left(S_{-\mu}\identity-\mathcal{O}_{\mu,i}\mathcal{O}^\ast_{\mu,i}\right),
\end{displaymath}
where the term
\begin{equation}\label{defn of S}
  S_\mu\identity:=\frac{1}{2L}\sum_k\left|\widehat{g}\left(\omega_k+\mu\right)\right|^2\identity
\end{equation}
arises from putting the individual operator products into normal order, using (\ref{anticommutation relations 2}).
Discarding the terms $\mathcal{O}_{\mu,i}^\ast \mathcal{O}_{\mu,i}$ and $\mathcal{O}_{\mu,i} \mathcal{O}_{\mu,i}^\ast$
which contribute positively to $T_f$, and combining the remaining integrals, the result is the state-independent lower
bound
\begin{equation}\label{the basic bound}
  \langle T_f\rangle_\psi \geq -\frac{1}{\pi}\int_0^\infty\difx{\mu}\mu S_\mu.
\end{equation}

\begin{figure}[t]
  \begin{center}
    \includegraphics{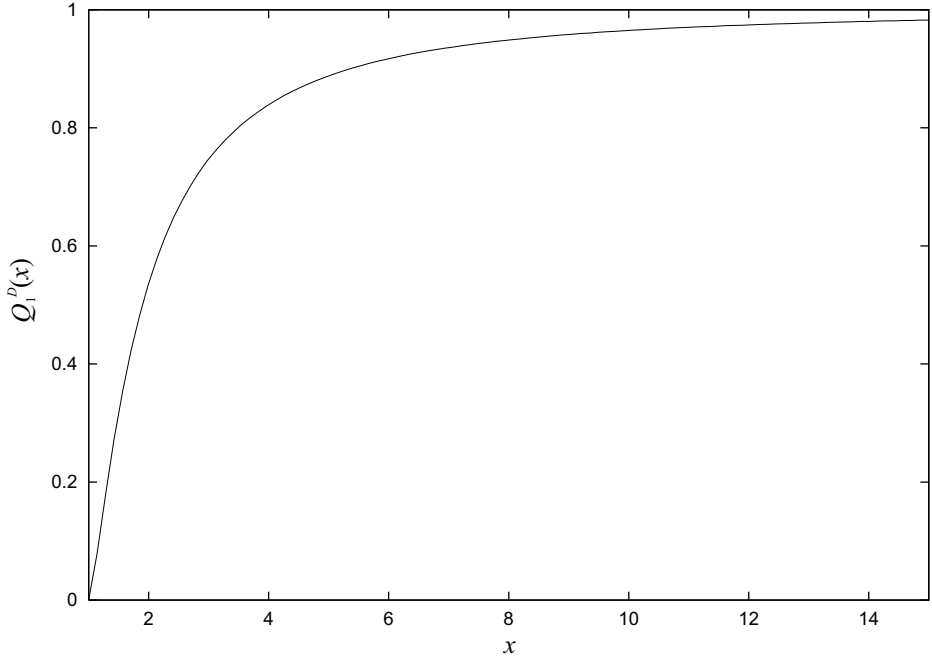}
    \caption{\label{graph of Q1}Plot of the function $Q_1^D$. Notice the asymptotic behaviour
                                $\lim_{x\rightarrow\infty}Q_1^D\left(x\right)=1$.}
  \end{center}
\end{figure}

It remains to calculate the right-hand side of (\ref{the basic bound}), and to show that it is finite. Substituting from
the definition (\ref{defn of S}) of $S_\mu$, and taking the continuum limit
\begin{displaymath}
  \frac{1}{2L}\sum_k\rightarrow\frac{1}{2\pi}\int_{-\infty}^\infty\difx{k},
\end{displaymath}
the final bound is found to be
\begin{displaymath}
  \langle T_f\rangle_\psi\geq-\frac{1}{2\pi^2}\int_m^\infty\difx{u}u^2\left|\widehat{g}\left(u\right)\right|^2
    Q_1^D\left(\frac{u}{m}\right),
\end{displaymath}
where the function $Q_1^D$ is defined by (\ref{definition of Q_1^D}). A graph of $Q_1^D$ is shown in figure \ref{graph
of Q1}; its qualitative similarity to the $Q_3^D$ of \cite{Fewster/Mistry} is easily seen. The scalar field analogue of
$Q_1^D$ is defined \cite{Fewster/Eveson} by
\begin{displaymath}
  Q_1\left(x\right):=\sqrt{1-\frac{1}{x^2}}+\frac{1}{x^2}\ln\left(x+\sqrt{x^2-1}\right),\quad 1\leq x\leq\infty.
\end{displaymath}
This is related to $Q_1^D$ by the identity
\begin{displaymath}
  Q_1^D\left(x\right)\equiv2\sqrt{1-\frac{1}{x^2}}-Q_1\left(x\right).
\end{displaymath}

Finally, invoking the translational invariance of the theory, the bound can be generalized to the worldline
$\left(t,x_0\right)$ of any stationary observer; this gives
\begin{displaymath}
  \int\difx{t}\langle:T_{00}\left(t,x_0\right):\rangle_\psi g\left(t\right)^2
    \geq-\frac{1}{2\pi^2}\int_m^\infty\difx{u}u^2\left|\widehat{g}\left(u\right)\right|^2
    Q_1^D\left(\frac{u}{m}\right).
\end{displaymath}
It only remains to show that the bound is finite (since it would otherwise be vacuous). Because $g\in
C^\infty_0\left(\mathbb{R}\right)$, it follows that $\left|\widehat{g}\left(u\right)\right|^2$ decays faster than any
inverse polynomial in $u$ as $u\rightarrow\infty$. But $u^2Q_1^D\left(u/m\right)$ grows like $u^2$ for large $u$, and so
the bound is finite.

\section{Massless limit}

The asymptotic behaviour $\lim_{x\rightarrow\infty}Q_1^D\left(x\right)=1$ leads immediately to the massless QWEI in the
form
\begin{displaymath}
  \langle T_f\rangle_\psi\geq-\frac{1}{2\pi^2}\int_0^\infty\difx{u}u^2\left|\widehat{g}\left(u\right)\right|^2.
\end{displaymath}
Again following the method of \cite{Fewster/Mistry}, the massless QWEI is cast into its final form:
\begin{equation}\label{zero mass limit}
  \langle T_f\rangle_\psi\geq-\frac{1}{2\pi}
    \int_{-\infty}^\infty\difx{t}\left|\frac{\rmd g}{\rmd t}\right|^2.
\end{equation}

This massless bound is weaker by a factor of 3 than Vollick's optimal bound \cite{Vollick}. When techniques analogous to
those used here are applied to the scalar field, as in \cite{Fewster/Eveson, Fewster/Teo}, they lead to a massless bound
that overestimates the optimal bound of Flanagan \cite{Flanagan} by a factor of $3/2$. As was mentioned earlier, the
essential difference between the present work and the derivation of the scalar field QWEIs of \cite{Fewster/Eveson,
Fewster/Teo} is that the latter use the convolution theorem in an essential way; here, this is replaced by a different
identity for Fourier transforms
--- specifically, (2.17) of \cite{Fewster/Mistry}. It seems likely that this difference accounts for the
`extra' factor of two found here. In any case: while the bound (\ref{the final bound}) is not optimal, it does have the
same order of magnitude as the optimal bound, in the massless limit.

It is worth remarking that (\ref{zero mass limit}) holds also for Dirac fields with non-zero mass; because $Q_1^D$ is
bounded between 0 and 1 (see figure \ref{graph of Q1}), the QWEI is at its least restrictive in the zero-mass limit.

\section{Conclusion}

An explicit lower bound has been derived, constraining the smeared energy density of the Dirac field in flat
two-dimensional spacetime. Unlike the earlier, optimal bound of Flanagan and Vollick, the bound holds for all
(non-negative) values of the field mass $m$. In the massless limit, comparison with the Flanagan-Vollick optimal bound
revealed that the bound given here overestimates the optimal bound by a factor of 3.

It seems reasonable to presume that the general method employed here can be extended to non-flat spacetimes; this is
indeed the case, and the results of this generalization will be reported elsewhere.

\section*{Acknowledgements}

The author wishes to thank Christopher Fewster, for his insightful contributions to our many discussions of negative
energy densities.

\end{document}